\documentclass[11pt]{article}

\begin{document}

\title{{\bf Notes on anelastic effects and thermal noise in
suspensions of test masses in interferometric gravitational-wave
detectors}}
\author{Constantin Brif \\
{\small\it LIGO Project, California Institute of Technology,
Pasadena, CA 91125} }
\date{}
\maketitle

\tableofcontents

\thispagestyle{empty}

\newpage

\section{Introduction}

The thermal noise is expected to be one of the main limiting factors 
on the sensitivity of interferometric gravitational-wave detectors 
like LIGO and VIRGO. Thermal fluctuations of internal modes of the
interferometer's test masses and of suspension modes will dominate
the noise spectrum at the important frequency range between 50 and 
200 Hz (seismic noise and photon shot noise dominate for lower
and higher frequencies, respectively). It is important to note that
off-resonant thermal noise level in high-quality systems is so low 
that it is unobservable in table-top experiments. 
Therefore, predictions of the thermal-noise spectrum in LIGO are 
based on a combination of theoretical models 
(with the fluctuation-dissipation theorem of statistical mechanics 
serving as a basis) and experimental measurements of quality factors 
of systems and materials involved.
It is assumed that losses in the test masses and suspensions will 
occur mainly due to internal friction in their materials, which is 
related to anelasticity effects in solids.

These informal notes comprise some basic results on the theory of
anelasticity and thermal noise in pendulum suspensions. This
collection is by no means complete and focus on aspects which are
of interest for the author. The original results can be found in
a number of books, research papers, and theses. Some of these sources 
are listed in a short bibliography at the end of the present text; 
a list of research papers (since 1990) devoted to various aspects of 
the thermal noise in interferometric gravitational-wave detectors was 
prepared by the author and is available as a separate document.

\section{Fluctuation-dissipation theorem}
\setcounter{equation}{0}

Consider a linear one-dimensional mechanical system with coordinate
$x(t)$. If a force $F(t)$ acts on the system, than in the frequency 
domain the force and the coordinate are related by 
\begin{equation}
x(\omega) = H(\omega) F(\omega) ,
\end{equation}
where $H(\omega)$ is the system response function. Then the spectral 
densities (see Appendix) are related by
\begin{equation}
S_x(\omega) = |H(\omega)|^2 S_F(\omega) .
\end{equation}
The impedance of the system is defined as 
$Z(\omega) = F(\omega)/v(\omega) = F(\omega)/[ i \omega x(\omega)]$.
Therefore, the impedance and the response function are related by
$Z(\omega) = 1/[ i \omega H(\omega)]$.

If the system is in equilibrium with the thermal bath of temperature
${\cal T}$, then the \emph{fluctuation-dissipation theorem} (FDT)
says that the spectral density of the thermal force is
\begin{equation}
  \label{eq:fdt}
S_F^{{\rm th}} (\omega) = 4 k_B {\cal T} {\rm Re} Z(\omega) ,
\end{equation}
where $k_B$ is the Boltzmann constant. The form (\ref{eq:fdt}) of
the FDT is valid in the classical regime, when the thermal energy
$k_B {\cal T}$ is much larger than the energy quantum $\hbar \omega$.
Using the FDT, one readily obtains the thermal noise spectrum
\begin{equation}
  \label{eq:fdt-x}
S_x^{{\rm th}}(\omega) = \frac{ 4 k_B {\cal T} }{\omega^2} 
{\rm Re} Y(\omega) ,
\end{equation}
where $Y(\omega) = 1/Z(\omega)$ is the admittance and
${\rm Re} Y(\omega) = {\rm Re} Z(\omega)/|Z(\omega)|^2$ is the
conductance. The FDT is the basis for calculations of the thermal 
noise spectrum in interferometric gravitational-wave detectors.

\subsection{Example: Damped harmonic oscillator}

Consider a damping harmonic oscillator of mass $m$, spring constant 
$k$, and damping constant $\gamma$. The equation of motion is
\begin{equation}
m \ddot{x} + \gamma \dot{x} + k x = F(t) .
\end{equation}
In the frequency domain this can be written as
\begin{equation}
(-m \omega^2 + i \gamma \omega + k) x(\omega) = F(\omega) .
\end{equation}
The impedance of this system is 
$Z(\omega) = \gamma + i (m \omega - k/\omega)$.
Then the FDT gives the spectrum densities:
\begin{equation}
\label{eq:dho-ns}
S_F^{{\rm th}} (\omega) = 4 k_B {\cal T} \gamma , \hspace{12mm}
S_x^{{\rm th}}(\omega) = \frac{ 4 k_B {\cal T} \gamma }{
(m \omega^2 - k)^2 + \gamma^2 \omega^2 } . 
\end{equation}

\section{Anelasticity of solids}
\setcounter{equation}{0}

The FDT means that if a system has no dissipation channel, thermal
fluctuations will be zero. For an ideal elastic spring without
friction, ${\rm Re} Z(\omega) = 0$, and there are no fluctuations:
$S_x^{{\rm th}}(\omega) = 0$. Deviations of solids from the ideal
elastic behavior (anelasticity) will result in internal friction
(dissipation) and related thermal noise. For gravitational-wave
detectors like LIGO, the test masses will be highly isolated, so 
the internal friction in materials of which the masses and their 
suspensions are made is believed to be the main source of 
dissipation and thermal noise.

\subsection{The complex Young modulus and the loss function}

Deformations of solids are usually described in terms of stress
$\sigma$ and strain $\epsilon$ (equivalents of mechanical restoring 
spring force $F_s$ and displacement $x$, respectively). Perfect 
elastic solids satisfy Hooke's law
\begin{eqnarray}
\sigma(t) = E \epsilon(t) ,
\end{eqnarray}
where $E$ is the (constant) Young modulus (an equivalent of the
spring constant $k$). Anelasticity can be described by introducing
the complex Young modulus (or the complex spring constant in a 
mechanical model). This is done in the frequency domain:
\begin{equation}
\label{eq:39}
E(\omega) = \frac{ \sigma(\omega) }{ \epsilon(\omega) } ,
\hspace{12mm} k(\omega) = \frac{ F_s(\omega) }{ x(\omega) } .
\end{equation}
If an external force $F(t)$ acts on a point mass $m$ attached to such
an anelastic spring, then the equation of motion in the frequency
domain is
\begin{equation}
[ -m \omega^2 + k(\omega)] x(\omega) = F(\omega) .
\end{equation}
The impedance of this system is
\begin{equation}
Z(\omega) = \frac{-m \omega^2 + k (\omega)}{ i \omega} ,
\end{equation}
and ${\rm Re}\, Z(\omega) = (1/\omega) {\rm Im}\, k(\omega)$.
Now, the FDT theorem gives the thermal noise spectrum:
\begin{equation}
\label{eq:anel-ns}
S_x^{{\rm th}}(\omega) = \frac{ 4 k_B {\cal T} }{ k_R \omega}
\frac{\phi(\omega)}{ (1- m \omega^2 /k_R)^2 + \phi^2 } . 
\end{equation}
Here, $k_R (\omega) \equiv {\rm Re}\, k(\omega)$, and
\begin{equation}
\phi(\omega) = \frac{ {\rm Im}\, k(\omega) }{{\rm Re}\, k(\omega) }
\end{equation}
is the so-called loss function. Note that $\phi = \tan \delta$, 
where $\delta$ is the angle by which strain lags behind stress.
The loss function $\phi$ is a measure of the energy dissipation
in the system. The rate at which energy is dissipated is 
$\overline{F_s \dot{x}}$. Then the energy dissipated per cycle by 
an anelastic spring is
\begin{equation}
\Delta {\cal E} = (2 \pi/\omega) \overline{F_s \dot{x}} .
\end{equation}
Taking $F_s = F_0 \cos \omega t$ and 
$x = x_0 \cos(\omega t - \delta)$, one finds
\begin{equation}
\Delta {\cal E} = \pi x_0 F_0 \sin\delta .
\end{equation}
If $\delta$ is small than the total energy of spring vibration 
is ${\cal E} = \frac{1}{2} x_0 F_0$. Then for $\delta \ll 1$,
one obtains
\begin{equation}
\phi = \frac{ \Delta {\cal E} }{2 \pi {\cal E}} .
\end{equation}

For small $\phi$ (which is usually the case for the internal friction
in materials used in detectors like LIGO), it is customary to neglect
the frequency dependence of $k_R$. Then one can write
$k(\omega) = k [1 + i \phi(\omega)]$, where $k = m \omega_0^2$ is
a constant (and $\omega_0$ is the resonant frequency). Though this is
a good approximation for many practical reasons, in general $k_R$ must
be frequency-dependent because real and imaginary parts of $k(\omega)$
are related via the Kramers-Kronig relations.

\subsection{Simple models}

Here we consider some simple models of anelasticity in solids. Neither 
of them gives a full description of the behavior of a real material,
but nevertheless they are useful from the didactic point of view.

\subsubsection{Perfect elastic solid}

The mechanical model of perfect elastic solid is a lossless spring.
In this case $\sigma = E \epsilon$, so $\phi =0$. There is no
dissipation and no thermal noise.

\subsubsection{Maxwell solid}

The mechanical model of Maxwell solid is a lossless spring in 
series with a dashpot. The dashpot provides a source of viscous 
friction with $\sigma = \eta \dot{\epsilon}$. Then for Maxwell solid 
stress and strain are related by equation
\begin{equation}
\dot{\epsilon} = E^{-1} \dot{\sigma} + \eta^{-1} \sigma .
\end{equation}
This equation shows that for a constant strain, stress decays 
exponentially. On the other hand, for a constant stress, strain
increases linearly, which is a very wrong description for 
crystalline solids. Going to the frequency domain, one obtains
\begin{equation}
E(\omega) = \frac{ \sigma(\omega) }{ \epsilon(\omega) }
= \frac{ i \omega \eta E}{E + i \omega \eta}
= \frac{ \omega^2 \eta^2 E + i \omega \eta E^2 }{E^2 + 
\omega^2 \eta^2 } 
\end{equation}
and $\phi(\omega) = E/(\eta \omega)$.

\subsubsection{Voigt-Kelvin solid}

The mechanical model corresponding to Voigt-Kelvin anelastic 
solid consists of a lossless spring and a dashpot in parallel, 
which corresponds to a damped harmonic oscillator. The relation 
between stress and strain reads
\begin{equation}
\eta \dot{\epsilon} + E \epsilon = \sigma .
\end{equation}
For a constant stress $\sigma_0$, strain changes exponentially with 
the decay time $\eta/E$ from its initial value $\epsilon_0$ to the 
equilibrium (Hooke) value $\sigma_0 /E$.
For a constant strain, stress is also constant, like in Hooke's law.
This is a good description for materials like cork, but it is not 
suitable for metals. In the frequency domain, one has
\begin{equation}
E(\omega) = E + i \eta \omega , \hspace{12mm}
\phi(\omega) = (\eta/E) \omega .
\end{equation}
Substituting this $\phi$ into Eq.~(\ref{eq:anel-ns}), we find
\begin{equation}
S_x^{{\rm th}}(\omega) = \frac{ 4 k_B {\cal T} \eta }{
(m \omega^2 - E)^2 + \eta^2 \omega^2 } .
\end{equation}
This is the same as Eq.~(\ref{eq:dho-ns}) for a damped harmonic 
oscillator with $E \leftrightarrow k$ and 
$\eta \leftrightarrow \gamma$.

\subsubsection{Standard anelastic solid}

Though the model of standard anelastic solid (SAS) does not gives a
complete account of properties of real metals, it describes quite 
well basic mechanisms responsible for anelastic effects. In fact,
if a dissipation mechanism has characteristic relaxation times 
for strain upon a constant stress and for stress upon a constant 
strain, then the SAS model gives an adequate description. The 
corresponding mechanical model consists of a spring in parallel with 
a Maxwell unit (which is a spring in series with a dashpot). 
Let $E_1$ and $E_2$ be the Young moduli of the separate spring and 
of the spring in the Maxwell unit, respectively, and $\eta$ be the 
dashpot constant, as usual. Then stress and strain are related by the 
following equation:
\begin{equation}
\label{eq:sas1}
\frac{E_2}{\eta} \sigma + \dot{\sigma} = 
\frac{E_1 E_2}{\eta} \epsilon + (E_1 + E_2) \dot{\epsilon} .
\end{equation}
For a constant strain $\epsilon_0$, stress decays exponentially
from its initial value $\sigma_0$ to the equilibrium (Hooke)
value $E_1 \epsilon_0$:
\begin{equation}
\sigma(t) = E_1 \epsilon_0 + ( \sigma_0 - E_1 \epsilon_0 )
e^{-t/\tau_{\epsilon}} , \hspace{12mm}
\tau_{\epsilon} = \eta/E_2 .
\end{equation}
Analogously, for a constant stress $\sigma_0$, strain decays 
exponentially from its initial value $\epsilon_0$ to the 
equilibrium (Hooke) value $\sigma_0 /E_1$:
\begin{equation}
\epsilon(t) = \frac{\sigma_0}{E_1} + \left( \epsilon_0 -
\frac{\sigma_0}{E_1} \right) e^{-t/\tau_{\sigma}} , \hspace{12mm}
\tau_{\sigma} = \frac{E_1 + E_2}{E_1 E_2} \eta .
\end{equation}
Then Eq.~(\ref{eq:sas1}) can be rewritten in the following form
\begin{equation}
\label{eq:sas2}
\sigma + \tau_{\epsilon} \dot{\sigma} = 
E_R ( \epsilon + \tau_{\sigma} \dot{\epsilon} ) ,
\end{equation}
where $E_R \equiv E_1$ is called the relaxed Young modulus.
Transforming to the frequency domain, one obtains
\begin{equation}
(1 + i \omega \tau_{\epsilon}) \sigma(\omega) =
E_R (1 + i \omega \tau_{\sigma}) \epsilon(\omega) .
\end{equation}
Then the complex Young modulus is given by
\begin{equation}
\label{eq:327}
E(\omega) = E_R \frac{1 + i \omega \tau_{\sigma}}{
1 + i \omega \tau_{\epsilon}} =
E_R \frac{(1 + \omega^2 \tau_{\sigma} \tau_{\epsilon})
+ i \omega (\tau_{\sigma} - \tau_{\epsilon})}{
1 + \omega^2  \tau_{\epsilon}^2 } .
\end{equation}
It is easy to see that
\begin{equation}
E(\omega) \approx \left\{ \begin{array}{ll}
E_R , & \ \ \ \ \omega \ll 1 \\
E_U , & \ \ \ \ \omega \gg 1 , \end{array} \right.
\end{equation}
where $E_U \equiv E_1 + E_2$ is called the unrelaxed Young modulus.
The loss function has the form
\begin{equation}
\phi(\omega) = \frac{\omega (\tau_{\sigma} - \tau_{\epsilon}) }{
1 + \omega^2 \tau_{\sigma} \tau_{\epsilon} } =
\Delta \frac{ \omega \bar{\tau} }{ 1 + \omega^2 \bar{\tau}^2 } ,
\end{equation}
where
\begin{equation}
\bar{\tau} = \sqrt{\tau_{\sigma} \tau_{\epsilon}} = \tau_{\epsilon}
\sqrt{\frac{E_U}{E_R}} , \hspace{12mm}
\Delta = \frac{E_U - E_R}{\sqrt{E_U E_R}} = 
\frac{\tau_{\sigma} - \tau_{\epsilon}}{
\sqrt{\tau_{\sigma} \tau_{\epsilon}} } .
\end{equation}
One sees that $\phi \propto \omega$ for $\omega \bar{\tau} \ll 1$
and $\phi \propto \omega^{-1}$ for $\omega \bar{\tau} \gg 1$.
The loss function has its maximum $\phi = \Delta/2$ at 
$\omega = \bar{\tau}^{-1}$. This is called the Debye peak.
This behavior is characteristic for processes with exponential
relaxation of stress and strain. $\bar{\tau}$ is the characteristic 
relaxation time and $\Delta$ is the relaxation strength.

\paragraph{Thermoelastic damping mechanism}

Zener pointed out that the SAS model with
\begin{equation}
\phi(\omega) = \Delta \frac{ \omega \bar{\tau} }{ 
1 + \omega^2 \bar{\tau}^2 } ,
\end{equation}
is suitable for describing processes in which the relaxation of 
stress and strain is related to a diffusion process.
One example of such a process is the so-called thermoelastic
damping. Consider a specimen which is a subject to a deformation
in such a way that one part of it expands and the other contracts
(e.g., a wire of a pendulum which bends near the top while the
pendulum swings). The temperature increases in the contracted part
and decreases in the expanded part. The resulting thermal diffusion 
leads to the dissipation of energy. This anelastic effect can be
described by the SAS model with the thermal relaxation strength
and relaxation time given by
\begin{equation}
\Delta = \frac{E_U {\cal T} \alpha^2}{C_v} , \hspace{12mm}
\bar{\tau} \simeq \frac{d^2}{D} ,
\end{equation}
where ${\cal T}$ is the temperature, $\alpha$ is the linear thermal 
expansion coefficient, $C_v$ is the specific heat per unit volume, 
$d$ is the characteristic distance heat must flow, and $D$ is the
thermal diffusion coefficient, $D = \varrho/C_v$, where $\varrho$
is the thermal conductivity. For a cylindrical wire of diameter
$d$, the frequency of the Debye peak is
\begin{equation}
\bar{f} = \frac{1}{2 \pi \bar{\tau}} \simeq 2.6 \frac{D}{d^2} .
\end{equation}

\subsection{Boltzmann's superposition principle}

While the SAS has certain general features in common with actual 
solids, it does not reproduce precisely the behavior of any real 
metal. Simple models considered above can be generalized by a
theory which only assumes that the relation between stress 
and strain is linear. This assumption was expressed by Boltzmann
in the form of a superposition principle: If the deformation 
$x_1 (t)$ was produced by the force $F_1 (t)$ and the deformation 
$x_2 (t)$ was produced by the force $F_2 (t)$, then the force
$F_1 (t) + F_2 (t)$ will produce the deformation 
$x_1 (t) + x_2 (t)$. On the other hand, the deformation can be
regarded as the independent variable. In this case the 
superposition principle states: If the force $F_1 (t)$ is
required to produce the deformation $x_1 (t)$ and the force 
$F_2 (t)$ is required to produce the deformation $x_2 (t)$,
then the force $F_1 (t) + F_2 (t)$ will be required to produce 
the deformation $x_1 (t) + x_2 (t)$.

Let us introduce the quantity $\lambda(t)$ which is called the
creep function and is the deformation resulting from the sudden
application at $t=0$ of a constant force of magnitude unity.
During an infinitesimal interval from $t$ to $t + d t$, the
applied force $F(t)$ can be approximated by a constant force
of magnitude $\dot{F} d t$. Then the superposition principle
gives
\begin{equation}
\label{eq:334}
x(t) = \int_{-\infty}^{t} \lambda(t-t') \dot{F}(t') d t' .
\end{equation}
Conversely, we may regard the deformation as a specified
function of time. Let us define the quantity $\kappa(t)$ which is 
called the stress function and is the force which must be applied 
in order to produce the step-function deformation 
$x(t) = \Theta(t)$ (here $\Theta(t)$ is 1 for $t \geq 0$ and 0
for $t<0$). Then the linear relationship is
\begin{equation}
\label{eq:335}
F(t) = \int_{-\infty}^{t} \kappa(t-t') \dot{x}(t') d t' .
\end{equation}
The relation between the creep function and the strain function
is rather complicated; in general they satisfy the following 
inequality
\begin{equation}
\lambda(t) \kappa(t) \leq 1 .
\end{equation}
For constant $\kappa(t) = k$ we recover Hooke's law
$F(t) = k x(t)$ and then $\lambda(t) = k^{-1}$.
Integrating by parts in Eq.~(\ref{eq:335}), we obtain another
expression of the superposition principle,
\begin{equation}
\label{eq:337}
F(t) = \int_{-\infty}^{t} f(t-t') x(t') d t' ,
\end{equation}
where
\begin{equation}
f(t) = \kappa(0) \delta(t) + \dot{\kappa}(t) .  
\end{equation}

The relationship between the force and the deformation becomes
very simple in the frequency domain. Toward this end we introduce
the functions 
\begin{equation}
f_p (t) = f(t) \Theta(t) , \hspace{12mm}
\kappa_p (t) = \kappa(t) \Theta(t) , \hspace{12mm}
\lambda_p (t) = \lambda(t) \Theta(t) , 
\end{equation}
which are zero for $t < 0$. Using these functions, one can expand 
the upper integration limit in Eqs.~(\ref{eq:334}), (\ref{eq:335}), 
and (\ref{eq:337}) to $\infty$. Then we just can use the fact that
a convolution in the time domain is a product in the frequency
domain. This gives
\begin{equation}
F(\omega) = i \omega \kappa_p (\omega) x(\omega)
= f_p (\omega) x(\omega) .
\end{equation}
Thus the Fourier transform of the stress function is simply related
to the complex spring constant of Eq.~(\ref{eq:39}):
\begin{equation}
k(\omega) = f_p (\omega) = i \omega \kappa_p (\omega) .
\end{equation}

\subsubsection{Example: Standard anelastic solid}

For the SAS the stress function is given by
\begin{equation}
\kappa(t) = E_R + (E_U - E_R) e^{-t/\tau_{\epsilon}} .
\end{equation}
It is straightforward to see that this function leads to the
first-order differential equation of the form (\ref{eq:sas2}).
Then we find the function $f(t)$,
\begin{equation}
f(t) = E_U \delta(t) - \frac{E_U - E_R}{\tau_{\epsilon}}
e^{-t/\tau_{\epsilon}} ,
\end{equation}
and the complex string constant,
\begin{equation}
k(\omega) = \int_{0}^{\infty} f(t) e^{- i \omega t} d t = 
E_U - \frac{E_U - E_R}{1 + i \omega \tau_{\epsilon}} .
\end{equation}
This can be rewritten in the form
\begin{equation}
k(\omega) = E_R \frac{1 + i \omega \tau_{\sigma}}{
1 + i \omega \tau_{\epsilon}} =
E_R \frac{(1 + \omega^2 \tau_{\sigma} \tau_{\epsilon})
+ i \omega (\tau_{\sigma} - \tau_{\epsilon})}{
1 + \omega^2  \tau_{\epsilon}^2 } .
\end{equation}
which coincides with Eq.~(\ref{eq:327}).

\section{Calculation of the thermal noise spectrum for a pendulum 
suspension}
\setcounter{equation}{0}

For a point mass $m$ attached to an anelastic spring with the
complex spring constant $k(\omega)$, we found a simple result
\[
Z(\omega) = \frac{k(\omega)-m \omega^2}{ i \omega} ,
\]
which can be used in the FDT to derive the thermal noise spectrum
$S_x^{{\rm th}}(\omega)$ as given by Eq.~(\ref{eq:anel-ns}). 
However, the question is how to find the thermal noise spectrum for 
more complicated systems, e.g., for pendulum suspensions of test 
masses in interferometric gravitational-wave detectors like LIGO.

In the literature we can find two different approaches: the 
``direct'' application of the FDT to the whole system and the
method which is based on decomposing a complicated system into
a set of normal modes. Below, we describe briefly both of these
approaches.

\subsection{The direct approach}

In brief, this method can be described as follows. First, one
should write equations of motion for the whole system and find
the impedance $Z(\omega)$. Then the FDT provides the thermal 
noise spectrum:
\begin{equation}
S_x^{{\rm th}}(\omega) = \frac{4 k_B T}{\omega^2} {\rm Re}\,
[ 1/Z(\omega) ] .
\end{equation}
The impedance $Z(\omega)$ contains the information about 
resonances of the system. The dissipation enters by taking
the Young moduli of the materials to be complex:
\begin{equation}
E(\omega) = [ 1 + i \phi(\omega)]\, {\rm Re}\, E(\omega) ,
\end{equation}
or, for simplicity, $E(\omega) = E_0 [ 1 + i \phi(\omega)]$,
where $E_0$ is a constant. The loss function $\phi(\omega)$
is obtained from experiments on the anelasticity of materials
used in the system (e.g., on the suspension wires).
Of course, the resulting noise spectrum $S_x^{{\rm th}}(\omega)$ 
depends very much on what form of $\phi(\omega)$ is used.

\subsection{The normal-mode decomposition}

The normal-mode decomposition is a more traditional approach.  
Consider, for example, an one-dimensional system of linear mass
density $\rho (z)$ and total length $L$, which is described in 
terms of the normal modes $\psi_n (z)$. These modes satisfy the 
orthonormality relation,
\begin{equation}
\int_{0}^{L} \rho (z) \psi_m (z) \psi_n (z) d z = \delta_{m n} , 
\end{equation}
and an arbitrary displacement $x(z,t)$ can be decomposed as
\begin{equation}
\label{eq:44}
x(z,t) = \sum_{n} \psi_n (z) q_n (t) .
\end{equation}
Here, $q_n (t)$ are the mode coordinates which satisfy
\begin{equation}
\ddot{q}_n + \omega_n^2 q_n = F_n (t) ,
\end{equation}
where $\omega_n$ are the resonance frequencies of the modes, and
\begin{equation}
F_n (t) = \int_{0}^{L} f(z,t) \psi_n (z) d z
\end{equation}
is the generalized force produced by the force density
$f(z,t)$ applied to the system.

This decomposition effectively replaces the complicated system by
a collection of oscillators, and each of them satisfies
\begin{equation}
[-\omega^2 + \omega_n^2 (\omega) ] q_n (\omega) = F_n (\omega) .
\end{equation}
The dissipation is included by taking
\begin{equation}
\omega_n^2 (\omega) = \omega_n^2 [ 1 + i \phi_n (\omega)] ,
\end{equation}
where $\phi_n (\omega)$ are the loss functions. Then we can write
\begin{equation}
\label{eq:49}
q_n (\omega) = \frac{ F_n (\omega) }{-\omega^2 + \omega_n^2 + 
i \omega_n^2 \phi_n (\omega)} .
\end{equation}

Let us assume that the force is applied at the end of the system
$z=L$, such that $f(z,t) = F(t) \delta(z-L)$. Then the generalized 
forces are $F_n (t) = F(t) \psi_n (L)$, and we can substitute
Eq.~(\ref{eq:49}) into the Fourier transform of Eq.~(\ref{eq:44}) 
to obtain
\begin{equation}
x(L,\omega) \equiv x(\omega) = 
\sum_n \frac{\psi_n^2 (L)}{-\omega^2 + \omega_n^2 + 
i \omega_n^2 \phi_n (\omega)} F(\omega) .
\end{equation}
This gives the admittance of the system in the form
\begin{equation}
Y(\omega) = 1/Z(\omega) = \sum_n \frac{i \omega \psi_n^2 (L)}{
-\omega^2 + \omega_n^2 + i \omega_n^2 \phi_n (\omega)} .
\end{equation}
Then the FDT can be used to obtain the spectral density of 
thermal fluctuations at $z=L$:
\begin{equation}
S_x^{{\rm th}}(\omega) = \frac{4 k_B {\cal T}}{\omega} \sum_n 
\frac{ \psi_n^2 (L) \omega_n^2 \phi_n (\omega) }{
( \omega_n^2 - \omega^2 )^2 + \omega_n^4 \phi_n^2 } .
\end{equation}
This can be written as a sum
\begin{equation}
S_x^{{\rm th}}(\omega) = \sum_n S_n^{{\rm th}} (\omega)
\end{equation}
over the contributions 
\begin{equation}
\label{eq:414}
S_n^{{\rm th}}(\omega) = \frac{ 4 k_B {\cal T} }{\omega}
\frac{k_n^{-1} \phi_n(\omega)}{ 
(1 - m_n \omega^2 /k_n)^2 + \phi_n^2 } =
\frac{ 4 k_B {\cal T} }{ \omega}
\frac{m_n^{-1} \omega_n^2 \phi_n(\omega)}{ 
( \omega_n^2 - \omega^2 )^2 + \omega_n^4 \phi_n^2 }
\end{equation}
of independent oscillators labeled by the index $n$. Each of these 
oscillators consists of a point mass $m_n = [\psi_n (L)]^{-2}$ 
attached to an anelastic spring with the complex spring constant 
$k_n (\omega) = k_n [ 1 + i \phi_n (\omega)]$, such that the
resonant angular frequencies are $\omega_n = \sqrt{k_n / m_n}$.
So, in order to obtain the thermal noise spectrum one needs to
find all the normal modes, their effective masses, resonant  
frequencies, and loss functions.

\subsection{Modes of a pendulum suspension}

The most important modes of a pendulum suspension are the pendulum
mode, the rocking mode, and the violin modes. We will not consider
here the rocking mode because for multi-loop suspensions the rocking
motion of the test mass is essentially suppressed. The loss function 
of each mode depends on the type of mode and on anelastic properties 
of the pendulum wire.  

\subsubsection{The pendulum mode}

For the pendulum mode, we will assume that the mass of the wire is
much smaller than the mass of the bob (which is the test mass) and 
that the bob is attached near its center of mass. Also, the angle
by which the pendulum swings is considered to be very small.
Then the pendulum may be modelled as an oscillator of the resonant
angular frequency
\begin{equation}
\omega_{{\rm p}} = \sqrt{g/L} ,
\end{equation}
where $g$ is the acceleration due to the Earth gravity field, and
$L$ is the pendulum length. 

The energy of the pendulum consists of two parts: the gravitational 
energy ${\cal E}_{{\rm gr}}$ and the elastic energy 
${\cal E}_{{\rm el}}$ due to the bending of the wire. The
gravitational energy is lossless; provided that all the losses
due to interactions with the external world (friction in the residual
gas, dumping by eddy currents, recoil losses into the seismic
isolation system, friction in the suspension clamps, etc.) are made 
insignificant by careful experimental design, the assumption is 
made that the losses are dominated by internal friction in the 
wire material. Consequently, 
$\Delta {\cal E} = \Delta {\cal E}_{{\rm el}}$.
Usually, ${\cal E}_{{\rm gr}} \gg {\cal E}_{{\rm el}}$, so we 
obtain for the pendulum-mode loss function:
\begin{equation}
\phi_{{\rm p}} = \frac{\Delta {\cal E}}{2 \pi 
{\cal E}_{{\rm tot}} }
= \frac{ \Delta {\cal E}_{{\rm el}} }{ 2 \pi 
({\cal E}_{{\rm el}} + {\cal E}_{{\rm gr}}) }
\approx \frac{ \Delta {\cal E}_{{\rm el}} }{ 2 \pi 
{\cal E}_{{\rm el}} } 
\frac{ {\cal E}_{{\rm el}} }{ {\cal E}_{{\rm gr}} } .
\end{equation}
Note that
\begin{equation}
\phi_{{\rm w}} = \frac{ \Delta {\cal E}_{{\rm el}} }{ 2 \pi 
{\cal E}_{{\rm el}} } 
\end{equation}
is the loss function for the wire itself which occurs due to
anelastic effects in the wire material. Then we obtain
\begin{equation}
\phi_{{\rm p}} = \xi_{{\rm p}} \phi_{{\rm w}} ,
\end{equation}
where $\xi_{{\rm p}}$ is the ratio between the elastic energy 
and the gravitational energy for the pendulum mode,
\begin{equation}
\xi_{{\rm p}} = \left( \frac{ {\cal E}_{{\rm el}} }{ 
{\cal E}_{{\rm gr}} } \right)_{{\rm p}} .
\end{equation}

The lossless gravitational energy of the pendulum is
\begin{equation}
\label{eq:420}
{\cal E}_{{\rm gr}} = 
\frac{1}{2} M \omega_{{\rm p}}^2 L^2 \theta_m^2
= \frac{1}{2} M g L \theta_m^2 ,
\end{equation}
where $M$ is the pendulum mass and $\theta_m$ is the maximum 
angle of swing.
The elastic energy depends on how many wires are used and how 
they are attached to the pendulum. For one wire, the fiber in 
the pendulum mode will bend mostly near the top, with the 
bending elastic energy
\begin{equation}
\label{eq:421}
{\cal E}_{{\rm el}} = \frac{1}{4} \sqrt{T E I} \theta_m^2 .
\end{equation}
Here, $T$ is the tension force in the wire ($T = M g$ for one 
wire), $E$ is the Young modulus of the wire material, and $I$
is the moment of inertia of the wire cross section 
($I = \frac{1}{2} \pi r^4$ for a cylindrical wire of radius $r$).
Using these results, one finds for a single-wire pendulum:
\begin{equation}
\xi_{{\rm p}} = \frac{ \sqrt{T E I} }{2 M g L} =
\frac{1}{2 L} \sqrt{ \frac{E I}{M g} } = 
\frac{1}{2 L} \sqrt{ \frac{E I}{T} } .
\end{equation}
This result can be easily generalized for the case when the test
muss is suspended by $N$ wires. Then the elastic energy
${\cal E}_{{\rm el}}$ of Eq.~(\ref{eq:421}) should be multiplied 
by $N$ and the tension in each wire becomes $T = M g/N$. Then
\begin{equation}
\label{eq:423}
\xi_{{\rm p}} = \frac{ N \sqrt{T E I} }{2 M g L}  =
\frac{1}{2 L} \sqrt{ \frac{E I N}{M g} } = 
\frac{1}{2 L} \sqrt{ \frac{E I}{T} } .
\end{equation}

In Eq.~(\ref{eq:423}) we assumed that all the wires are in one plane:
a plane through the center of mass of the pendulum, whose normal is
parallel to the direction of swing. (Note that in such an 
configuration one should take into account the rocking mode of the 
test mass.) In this arrangement, the pendulum mode causes bending of 
the wires mostly at the top. If one uses a number of wire loops along
the test mass length, then the rocking mode is essentially 
suppressed and the wires bend both at the top and the bottom.
Therefore, the bending elastic energy of the multi-loop configuration 
is given by multiplying the result of Eq.~(\ref{eq:421}) by $2 N$,
\begin{equation}
{\cal E}_{{\rm el}} = \frac{N}{2} \sqrt{T E I} \theta_m^2 .
\end{equation}
Then the energy ratio is
\begin{equation}
\xi_{{\rm p}} = \frac{ N \sqrt{T E I} }{M g L}  =
\frac{1}{L} \sqrt{ \frac{E I N}{M g} } = 
\frac{1}{L} \sqrt{ \frac{E I}{T} } .
\end{equation}

The contribution of the pendulum mode to the thermal noise
spectrum is obtained from Eq.~(\ref{eq:414}) by taking 
$m_n = M$, $k_n = M g/L$, $\omega_n = \omega_{{\rm p}}$ and 
$\phi_n = \phi_{{\rm p}} = \xi_{{\rm p}} \phi_{{\rm w}}$. 
This gives
\begin{equation}
S_{{\rm p}}^{{\rm th}}(\omega) = 
\frac{ 4 k_B {\cal T} }{\omega M}
\frac{\omega_{{\rm p}}^2 \phi_{{\rm p}} (\omega)}{
( \omega_{{\rm p}}^2 - \omega^2 )^2 
+ \omega_{{\rm p}}^4 \phi_{{\rm p}}^2 } .
\end{equation}
For LIGO suspensions, $f_{{\rm p}} = \omega_{{\rm p}}/2 \pi$ is 
about 1 Hz. This is much below the working frequency range  
(near 100 Hz), so we may assume 
$\omega_{{\rm p}} /\omega \ll 1$.
Also, the loss function is very small, 
$\phi_{{\rm p}} < 10^{-5}$.
Then the pendulum-mode contribution to the thermal noise
spectrum is
\begin{equation}
S_{{\rm p}}^{{\rm th}}(\omega) \simeq \frac{ 4 k_B {\cal T} 
\omega_{{\rm p}}^2 \phi_{{\rm p}} (\omega)}{M \omega^5}
= \frac{ 4 k_B {\cal T} }{ L^2 } \sqrt{ \frac{g E I N}{M^3} } 
\frac{\phi_{{\rm w}} (\omega)}{\omega^5} .
\end{equation}

\subsubsection{The violin modes}

The angular frequency of the $n$th violin mode ($n=1,2,3,\ldots$)
is given by
\begin{equation}
\omega_n = \frac{n \pi}{L} \sqrt{\frac{T}{\rho}} \left[
1 + \frac{2}{k_e L} + \frac{1}{2} \left( \frac{n \pi}{k_e L}
\right)^2 \right] ,
\end{equation}
where $L$ is the length of the wire, $T$ is the tension force,
$\rho$ is the linear mass density of the wire, and 
\begin{equation}
k_e \simeq \sqrt{\frac{T}{E I}} .
\end{equation}
In the violin mode the wire bends near both ends in a similar way.
The bending occurs over the characteristic distance scale 
$k_e^{-1} \simeq \sqrt{E I/T}$, the same as in the pendulum mode. 
For $k_e^{-1} \ll L$, which is a very good estimation for
heavily loaded thin wires like in LIGO, one have approximately,
\begin{equation}
\label{eq:430}
\omega_n \simeq \frac{n \pi}{L} \sqrt{\frac{T}{\rho}} .
\end{equation}
This is just the angular frequency of the $n$th vibrational mode 
of an ideal spring.

It can be shown that for the $n$th violin mode, the loss function is
\begin{equation}
\phi_n = \xi_n \phi_{{\rm w}} , \hspace{12mm}
\xi_n = \left( \frac{ {\cal E}_{{\rm el}} }{ {\cal E}_{{\rm gr}} }    
\right)_n ,
\end{equation}
where the energy ratio is
\begin{equation}
\label{eq:432}
\xi_n = \frac{2}{k_e L} \left( 1 + \frac{ n^2 \pi^2 }{2 k_e L}
\right) \simeq \frac{2}{L} \sqrt{\frac{E I}{T}} \left( 1 + 
\frac{1}{2 L} \sqrt{\frac{E I}{T}} n^2 \pi^2 \right) .
\end{equation}
Since $k_e L \gg 1$, for first several modes the energy ratio is 
approximately
\begin{equation}
\xi_n \simeq \xi_{{\rm v}} = \frac{2}{L} \sqrt{\frac{E I}{T}} .
\end{equation}
This expression takes into account only the contribution to the 
elastic energy due to wire bending near the top and the bottom. 
For higher violin modes, one should also consider the 
contribution due to wire bending along its length, which leads to 
Eq.~(\ref{eq:432}).

For the one-loop suspension configuration, the elastic energy
of the lowest violin modes is about twice of that for the
pendulum mode (for the last one the wires bend only at the top
while for the former ones the wires bend at both ends).
In the multi-loop configuration, the elastic energy of the lowest 
violin modes and of the pendulum mode is approximately the same.
On the other hand, the gravitational energy of the pendulum mode
is by a factor of 2 larger than that of a violin mode.
For the violin modes of each wire, the gravitational energy is
$\frac{1}{4} T L \theta_m^2$. Then for $N$ wires,
\begin{equation}
( {\cal E}_{{\rm gr}} )_{{\rm v}} = \frac{1}{4} N T L \theta_m^2
= \frac{1}{4} M g L \theta_m^2 .
\end{equation}
This is just one half of the gravitational energy for the 
pendulum mode, $( {\cal E}_{{\rm gr}} )_{{\rm p}} = 
\frac{1}{2} M g L \theta_m^2$ (cf. Eq.~(\ref{eq:420})). 
This explains the difference between the loss functions for the
pendulum mode and for the violin modes: 
$\xi_{{\rm v}} \simeq 4 \xi_{{\rm p}}$ 
for the one-loop configuration and 
$\xi_{{\rm v}} \simeq 2 \xi_{{\rm p}}$ 
for the multi-loop configuration.

The effective mass of the $n$th violin mode is
\begin{equation}
m_n = [\psi_n (L)]^{-2} = \frac{1}{2} N M 
\left( \frac{\omega_n}{\omega_{{\rm p}}} \right)^2
\simeq \frac{ \pi^2 M^2 }{2 \rho L} n^2 , 
\end{equation}
where we took expression (\ref{eq:430}) for $\omega_n$ and 
$T = M g/N$. This effective mass arises because the violin 
vibrations of the wire cause only a tiny recoil of the test 
mass $M$. The contribution of the violin modes to the thermal 
noise spectrum is given by
\begin{equation}
S_{{\rm v}}^{{\rm th}}(\omega) = \frac{4 k_B {\cal T}}{\omega} 
\sum_{n=1}^{\infty} \frac{ m_n^{-1} \omega_n^2 \phi_n (\omega) 
}{ ( \omega_n^2 - \omega^2 )^2 + \omega_n^4 \phi_n^2 } .
\end{equation}
Typical values of $f_1 = \omega_1 /2\pi$ are from 350 to 500 Hz.
If we are interested in the thermal spectrum density near 
100 Hz, we can assume $\omega^2 \ll \omega_n^2$. Then we have 
approximately
\begin{equation}
\label{eq:437}
S_{{\rm v}}^{{\rm th}}(\omega) \simeq \frac{ 8 k_B {\cal T} 
\omega_{{\rm p}}^2 }{ N M \omega} \sum_{n=1}^{\infty} 
\frac{ \phi_n (\omega) }{ \omega_n^4 } \simeq
\frac{ 8 k_B {\cal T} N \rho^2 L^3 }{ \pi^4 g M^3 \omega}
\sum_{n=1}^{\infty} \frac{ \phi_n (\omega) }{ n^4 } .
\end{equation}
One can see that the contributions of higher violin modes are
very small due to the factor $n^{-4}$ in the sum.
Taking $\phi_n = \xi_n \phi_{{\rm w}}$ and using 
Eq.~(\ref{eq:432}), we obtain
\begin{equation}
\label{eq:438}
\sum_{n=1}^{\infty} \frac{ \phi_n (\omega) }{ n^4 }
= \frac{2}{k_e L} \left( \frac{ \pi^4 }{90} + 
\frac{ \pi^4 }{12 k_e L} \right) \phi_{{\rm w}} (\omega) 
\simeq \frac{ \pi^4 }{45 L} \sqrt{ \frac{E I}{T} } 
\phi_{{\rm w}} (\omega) .
\end{equation}
Here, we assumed $k_e L \gg 1$. Finally, we substitute 
(\ref{eq:438}) into (\ref{eq:437}) and find the following
expression for the violin-mode contribution to the thermal
noise spectrum,
\begin{equation}
S_{{\rm v}}^{{\rm th}}(\omega) \simeq 
\frac{8}{45} k_B {\cal T} \rho^2 L^2  
\sqrt{ \frac{ E I N^3 }{ g^3 M^7 } } 
\frac{ \phi_{{\rm w}} (\omega) }{\omega} .
\end{equation}

\section{Experiments on anelasticity effects for pendulum 
suspensions}
\setcounter{equation}{0}

\subsection{Basic types of experiments}

In order to predict the thermal noise fluctuations in pendulum
suspensions, two basic types of experiments are performed:
\begin{enumerate}
\item Investigations of anelastic properties of wires made of
various materials, in order to determine the wire loss function
$\phi_{{\rm w}} (\omega)$.
\item Measurements of quality factors ($Q = \phi^{-1}$ at a
resonance) for the pendulum and violin modes of actual 
suspensions, in order to verify the relationships
\begin{equation}
\label{eq:51}
\phi_{{\rm p}} (\omega) = \xi_{{\rm p}} \phi_{{\rm w}} (\omega) , 
\hspace{12mm}
\phi_{{\rm v}} (\omega) = \xi_{{\rm v}} \phi_{{\rm w}} (\omega) .
\end{equation}
\end{enumerate}
Early experiments showed serious discrepancy between the measured
quality factors and those predicted using Eq.~(\ref{eq:51}).
It was discussed that this discrepancy may happen due to 
stress-dependent effects. However, it was shown later that the
internal losses of the wires are almost independent of the applied
stress. Many recent experiments proved that the above discrepancy
appears due to serious losses in the clamps. A smart design of 
clamps can be used to reduce these excess losses and then 
predictions of Eq.~(\ref{eq:51}) are quite accurate. A very
promising possibility is the use of monolithic or semi-monolithic
suspensions. The design of clamps plays a crucial role in the 
reduction of the thermal noise of the test mass suspensions.

\subsection{Internal losses in wire materials}

A number of experiments were performed to study internal losses
of various wire materials (e.g., steel, tungsten, 
fused quartz, and some others). The main drawback of many of
these experiments is a small number of frequencies for which
$\phi_{{\rm w}}$ was measured. Also, there are serious 
discrepancies between results of different experiments. 
Therefore, the exact frequency dependence of $\phi_{{\rm w}}$ 
is still unclear for many materials. Below, we briefly review 
results of some recent experiments.

\paragraph{Kovalik and Saulson, 1993}
Method: Quality factors were measured for resonances of freely 
suspended wires. Materials: Tungsten, silicon, sapphire, fused 
quartz. Results: Insignificant frequency dependence for tungsten;
for fused quartz, measured $\phi_{{\rm w}}$ are above those 
predicted by the thermoelastic damping (TED) for some frequencies 
and near TED for others; sapphire and silicon showed behavior 
consistent with TED.

\paragraph{Saulson et al., 1994}
Method: Quality factors were measured for an inverted pendulum of 
tunable length. Material: Free-Flex cross-spring flexure made of
crossed steel strips. Results: In agreement with a 
frequency-independent $\phi_{{\rm w}}$.

\paragraph{Gillespie and Raab, 1994}
Method: Quality factors were measured for resonances of freely 
suspended wires. Material: Steel music wires. Results: A constant 
value of $\phi_{{\rm w}}$ for low frequencies (from 30 to 150 Hz). 
For higher frequencies (from 150 Hz to 2 kHz) $\phi_{{\rm w}}$ 
increases with $\omega$, like TED predicts, but the measured value 
$\phi_{{\rm meas}}$ is well above $\phi_{{\rm TED}}$. 
These results may be explained by the formula 
$\phi_{{\rm meas}} = \phi_{{\rm TED}} + \phi_{{\rm ex}}$,
where $\phi_{{\rm ex}}$ is a frequency-independent excess loss.

\paragraph{Rowan et al., 1997}
Method: Quality factors were measured for resonances of ribbons 
fixed at one end. Material: Fused quartz ribbons. 
Results: Data were obtained for 5 resonances in the range from 
6 to 160 Hz. $\phi_{{\rm meas}}$ is well above $\phi_{{\rm TED}}$
for lower frequencies (below 30 Hz), and in agreement with TED
for higher frequencies (above 80 Hz).

\paragraph{Dawid and Kawamura, 1997}
Method: Quality factors were measured for the violin modes
of wires fixed at both ends in a ``guitar''-type apparatus.
Materials: Invar, titanium, steel, tungsten and several other
metals. Results: $\phi_{{\rm meas}}^{-1}$ was proportional to
$\sqrt{T}$, in accordance with the formula
$\phi_{{\rm v}} = (2/L) \sqrt{E I/ T} \phi_{{\rm w}}$ for 
frequency-independent $\phi_{{\rm w}}$.

\paragraph{Huang and Saulson, 1998}
Method: Quality factors were measured for resonances of freely 
suspended wires. Materials: Steel and tungsten. For steel, 
$\phi_{{\rm meas}}$ coincides with the predictions of TED
(the characteristic Debye-peak frequency dependence).
Some differences were found between properties of annealed
wires ($\phi_{{\rm meas}}$ slightly above $\phi_{{\rm TED}}$)
and ``curly'' wires ($\phi_{{\rm meas}}$ slightly below 
$\phi_{{\rm TED}}$). The difference can be explained by 
modifications of thermal properties. For tungsten wires, 
$\phi_{{\rm meas}}$ only slightly increases with frequency;
the loss function increases with the wire diameter, as
should be for TED at frequencies well below $\bar{f}$.

\section{Conclusions}

It is seen that predictions of the spectral density for thermal
fluctuations in pendulum suspensions depend strongly on the type
of the dissipation mechanism. Sources of external losses
(friction in the residual gas, dumping by eddy currents, recoil 
losses into the seismic isolation system, friction in the 
suspension clamps, etc.) should be eliminated by careful 
experimental design. In particular, results of many recent 
experiments show that excess losses in clamps may seriously 
deteriorate the quality factors of suspension resonances.
When external losses are made sufficiently small, the main source 
of dissipation is the internal friction in the wires due to 
anelastic effects. The thermal noise spectrum depends on the form 
of the loss function. Unfortunately, the exact frequency 
dependence of the wire loss function $\phi_{{\rm w}} (\omega)$ is 
not yet completely understood. In many experiments 
$\phi_{{\rm w}}$ was measured only at few frequencies and 
experimental uncertainty of results was often quite large. 
Moreover, there is a contradiction between results of different 
experiments. Therefore, it is very difficult to make certain 
conclusions about the behavior of $\phi_{{\rm w}} (\omega)$. 
In particular, it is unclear if clamp losses 
are negligible in experiments with freely suspended wires, as is 
usually assumed. Certainly, there is a room for more experiments 
on anelastic properties of wires, in order to clarify the issue
of internal friction in the frequency range of interest for
gravitational-wave detection.

\section*{Acknowledgment}

This work would be impossible without great help and encouragement  
by Malik Rakhmanov. I thank him for long hours of illuminating 
discussions and for encouraging me to enter the realm of thermal
noise and anelasticity.

\section*{Appendix: Correlation function and spectral density}
\addcontentsline{toc}{section}{Appendix: Correlation function 
and spectral density}

Consider a system characterized by some quantity $\alpha$ (e.g., 
position or velocity). For stationary processes, the correlation
function is
\[
\rho_{\alpha}(t) = \langle \alpha(\tau) \alpha(\tau + t) \rangle ,
\]
where the average is over a statistical ensemble. Using the ergodic
theorem, this can be replaced by the time average,
\[
\rho_{\alpha}(t) = \lim_{T \rightarrow \infty} \frac{1}{T} 
\int_{-T}^{T} \alpha(t') \alpha(t' + t) d t' .
\]
Now, define the function
\[
\alpha_T (t) = \left\{ \begin{array}{ll}
\alpha(t) , & \ \ \ \ t \in [-T,T] \\ 0 , & \ \ \ \ {\rm other} 
\end{array} \right.
\]
and its Fourier transform
\[
\alpha_T (\omega) = \int_{-\infty}^{\infty} \alpha_T (t)
e^{ i \omega t} d t .
\]
The definition of the spectral density is
\[
S_{\alpha}(\omega) = \lim_{T \rightarrow \infty} \frac{ 
|\alpha_T (\omega)|^2 }{\pi T} .
\]
It is easy to see that the correlation function $\rho_{\alpha}(t)$
and the spectral density $S_{\alpha}(\omega)$ are related via the
Fourier transform:
\[
\rho_{\alpha}(t) = \frac{1}{2} \int_{-\infty}^{\infty} 
S_{\alpha}(\omega) e^{- i \omega t} d \omega , \hspace{12mm}
S_{\alpha}(\omega) = \frac{1}{\pi} \int_{-\infty}^{\infty} 
\rho_{\alpha}(t) e^{ i \omega t} d t .
\]
This result is known as the Wiener-Khinchin theorem.

\section*{Bibliography}
\addcontentsline{toc}{section}{Bibliography}
\begin{flushleft}
The fluctuation-dissipation theorem was introduced in 

H. B. Callen and T. A. Welton, 
``Irreversibility and Generalized Noise,''
Phys. Rev. {\bf 83}, 34 (1951); 

H. B. Callen and R. F. Greene, 
``On a Theorem of Irreversible Thermodynamics,''
Phys. Rev. {\bf 86}, 702 (1952). 

The theorem is discussed in a number of textbooks on statistical
physics, for example, 

L. E. Reichl, \emph{A Modern Course in Statistical Physics}
(Univ. Texas Press, Austin, 1980); 

L. D. Landau and E. M. Lifshitz, \emph{Statistical Physics}
(Pergamon Press, Oxford, 1980). 
\vskip 3mm

The theory of anelasticity is discussed in 

C. Zener, \emph{Elasticity and Anelasticity of Metals}
(Univ. Chicago Press, Chicago, 1948); 

A. S. Novick and B. S. Berry, \emph{Anelastic Relaxation in
Crystalline Solids} (Academic Press, New York, 1972). 

The theory of thermoelastic damping was presented by Zener: 

C. Zener, ``Theory of Internal Friction in Reeds,''
Phys. Rev. {\bf 52}, 230 (1937); 

C. Zener, ``General Theory of Thermoelastic Internal Friction,''
Phys. Rev. {\bf 53}, 90 (1938).
\vskip 3mm

Thermal fluctuations of pendulum suspensions and related problems
were discussed in many works. Some of them are listed below: 

P. R. Saulson, ``Thermal noise in mechanical experiments,''
Phys. Rev. D {\bf 42}, 2437 (1990); 

G. I. Gonz\'{a}lez and P. R. Saulson, 
``Brownian motion of a mass suspended by an anelastic wire''
J. Acoust. Soc. Am. {\bf 96}, 207 (1994); 

J. E. Logan, J. Hough, and N. A. Robertson,
``Aspects of the thermal motion of a mass suspended as a 
pendulum by wires,''
Phys. Lett. A {\bf 183}, 145 (1993); 

A. Gillespie and F. Raab, ``Thermal noise in the test mass 
suspensions of a laser interferometer gravitational-wave 
detector prototype,''
Phys. Lett. A {\bf 178}, 357 (1993); 

J. Gao, L. Ju, and D. G. Blair, 
``Design of suspension systems for measurement of high-Q 
pendulums,''
Meas. Sci. Technol. {\bf 6}, 269 (1995);


V. B. Braginsky, V. P. Mitrofanov, and K. V. Tokmakov,
``On the thermal noise from the violin modes of the test 
mass suspension in gravitational-wave antennae,''
Phys. Lett. A {\bf 186}, 18 (1994); 

V. B. Braginsky, V. P. Mitrofanov, and K. V. Tokmakov,
``Energy dissipation in the pendulum mode of the test mass 
suspension of a gravitational wave antenna,''
Phys. Lett. A {\bf 218}, 164 (1996); 

G. Cagnoli, L. Gammaitoni, J. Kovalik, F. Marchesoni, and 
M. Punturo,
``Suspension losses in low-frequency mechanical pendulums,''
Phys. Lett. A {\bf 213}, 245 (1996).
\vskip 3mm

Experiments on internal friction in various types of wires 
(see Sec.~5.2) were reported in the following papers:

J. Kovalik and P. R. Saulson, 
``Mechanical loss in fibers for low-noise pendulums,'' 
Rev. Sci. Instrum. {\bf 64}, 2942 (1993);

P. R. Saulson, R. T. Stennins, F. D. Dumont, and S. E. Mock,
``The inverted pendulum as a probe of anelasticity,''
Rev. Sci. Instrum. {\bf 65}, 182 (1994);

A. Gillespie and F. Raab, ``Suspension losses in the pendula 
of laser interferometer gravitational-wave detectors,''
Phys. Lett. A {\bf 190}, 213 (1994);

S. Rowan, R. Hutchins, A. McLaren, N. A. Robertson, 
S. M. Twyford, and J. Hough,
``The quality factor of natural fused quartz ribbons over a 
frequency range from 6 to 160 Hz,'' 
Phys. Lett. A {\bf 227}, 153 (1997);

D. J. Dawid and S. Kawamura, ``Investigation of violin mode Q 
for wires of various materials,''
Rev. Sci. Instrum. {\bf 68}, 4600 (1997);

Y. L. Huang and P. R. Saulson, ``Dissipation mechanisms in 
pendulums and their implications for gravitational wave 
interferometers,''
Rev. Sci. Instrum. {\bf 69}, 544 (1998).
\end{flushleft}

\end{document}